\documentclass[pre,aps,floats,superscriptaddress,floatfix,twocolumn]{revtex4}
\usepackage{amssymb,amsmath}
\usepackage{amsmath,amssymb}
\usepackage{graphicx}
\usepackage{psfrag}
\usepackage{color}

\def\beq{\begin{equation}}
\def\eeq{\end{equation}}
\def\bea{\begin{eqnarray}}
\def\eea{\end{eqnarray}}
\begin{document}
\title{Symmetries and scaling in generalised coupled conserved 
Kardar-Parisi-Zhang equations}
 
 \author{Tirthankar Banerjee}\email{tirthankar.banerjee@saha.ac.in}
\author{Abhik Basu}\email{abhik.basu@saha.ac.in,abhik.123@gmail.com}
\affiliation{Condensed Matter Physics Division, Saha Institute of
Nuclear Physics, Calcutta 700064, India}

\date{\today}
\begin{abstract}
 
  We study the noisy nonequilibrium dynamics of a conserved 
density that is driven by a 
fluctuating surface governed by the conserved Kardar-Parisi-Zhang equation. We 
uncover the
universal scaling properties of the conserved density. We consider two separate 
minimal models where 
the surface fluctuations couple (i) with the spatial variation of the conserved 
density, and (ii) directly
with the magnitude of the conserved density. Both these two models conserve 
the density, but differ from symmetry stand point. We use our result to 
highlight the
dependence of 
nonequilibrium universality
classes on the interplay between symmetries and conservation laws.

\end{abstract}

\maketitle

\section{Introduction}

The concept of universality classes,
which
are parametrised by the space dimensions and the order parameter components, 
allows one to have a systematic physical understanding of universal scaling 
properties in equilibrium systems~\cite{fisher,chaikin}. These universality 
classes
are found to be robust against dynamical perturbations so long as the general
conditions for 
equilibrium are maintained.  
In contrast, statistical properties of truly nonequilibrium
dynamic phenomena in systems with generic non-Gibbsian 
distribution are
found to be strongly sensitive to all kinds of perturbations.
Prominent examples are driven diffusive systems~\cite{driven} and
diffusion-limited reactions~\cite{react-diff}. For instance, one finds
that for the Kardar-Parisi-Zhang (KPZ) equation of surface 
growth~\cite{kpz,stanley}, that shows paradigmatic nonequilibrium phase 
transitions~\cite{natter},  anisotropic
perturbations are relevant in $d > 2$ spatial dimensions,
leading to rich phenomena that include novel universality
classes and the possibility of first-order phase transitions
and multicritical behavior~\cite{kpz-aniso}.
Furthermore, novel nonequilibrium scaling behaviour including continuously 
varying universality classes are often found in 
multicomponent driven 
systems~\cite{abfrey}. Related physical realisations include  
driven 
symmetric mixture
of a miscible binary fluid~\cite{binfluid} and magnetohydrodynamic 
turbulence~\cite{3dmhd}, dynamic roughening
of strings moving in random media~\cite{ertas}, sedimenting
colloidal suspensions~\cite{colloid} and crystals~\cite{driv-cryst}.

Conservation laws are known to play significant roles in physical systems. For 
equilibrium systems, they affect only the dynamical properties~\cite{halpin}, 
where as for 
out of equilibrium systems even time independent quantities are 
affected by conservation laws. 
This was succinctly brought out by 
the studies on a conserved version of the KPZ equation (C-KPZ) that shows  
scaling behaviour distinctly different from the usual KPZ 
equation~\cite{ckpz}. For instance, the KPZ universality class is 
characterised by the {\em exact} relation between the scaling exponents: 
$\chi_{kpz} + z_{kpz}=2$~\cite{natter,stanley}, where $\chi_{kpz}$ and 
$z_{kpz}$, respectively, are the roughness and dynamic scaling exponents 
describing the spatial and temporal scaling of the KPZ universality class. 
In contrast, the C-KPZ equation does not admit any such exact exponent 
relations~\cite{ckpz,janssen}. The KPZ equation has subsequently been 
generalised to multicomponent versions to address different questions of 
principles. For example, how the surface fluctuations in the KPZ equation 
control the fluctuations of a conserved scalar density that is dynamically 
coupled to the KPZ equation has been studied~\cite{ab-jkb-mct} by using a 
well-known two-component variant of the KPZ equation~\cite{ertas}. It is 
also known that a breakdown of an external symmetry like parity can lead to 
novel scaling behaviour~\cite{abfrey}.
Notable previous works that form a major motivation for our studies here are 
the studies 
reported by Drossel and Kardar (hereafter DK) in 
Refs.~\cite{dro1,dro2} using a set of coupled generalised KPZ equations for the 
height field and a density. In particular, DK studied fluctuations in 
the concentrations of structureless particles advected by a 
one-dimensional ($1d$) Burger's fluid, or 
equivalently particles sliding on a fluctuating KPZ surface~\cite{dro1}. By 
retaining
feedback from the density fluctuations on the fluctuating KPZ surface, they 
elucidated various regimes depending upon the choice of parameters for advection 
or anti-advection. The scaling exponents are obtained. Remarkably, continuously 
varying scaling exponents are illustrated for the anti-advection case in 
Ref.~\cite{dro1}. In a subsequent study, DK considered the interplay between a 
fluctuating surface and phase ordering~\cite{dro2}, again using a set of 
coupled generalised KPZ equations for the height field and a nonconserved 
density. They obtained the relevant scaling exponents and in some cases 
illustrated continuously varying dynamic exponent in the model.  These studies 
by DK open 
up the questions: (i) How do
 the internal symmetries of the equations of motion that control 
the structure of the nonlinear dynamical cross-coupling terms between the 
different fields conspire 
with 
the conservation laws to determine the universal scaling behaviour? (ii) How 
does the 
conservation law for the surface fluctuations affect the dynamics and 
fluctuations of an attached density?

In order to 
systematically address these generic issues, we study how a conserved 
fluctuating 
surface described by the C-KPZ 
equation affects the spatio-temporal properties of a conserved scalar density 
that is 
dynamically coupled to the fluctuating surface. When there are multiple 
dynamically coupled fields, with all of them exhibiting dynamical scaling, it is not 
{\em apriori} clear whether or not they should all have the same dynamic 
exponent; in case of equal dynamic exponents the model is said to display {\em 
strong} dynamic scaling, else {\em weak} dynamic scaling 
ensues~\cite{strong-weak}. In a study on coupled one-dimensional model, 
Ref.~\cite{weak} showed 
the sensitive dependence of the nature of dynamic scaling on the precise forms 
of the dynamic couplings in the model equations. In a model with several 
dynamical fields, one must thus distinguish between strong and weak 
dynamic scaling. These theoretical issues form 
the major motivation of the present work. Independent of any specific 
applications, the general importance of 
our studies here lie in their ability to identify ingredients that may control 
long-time, large-distance universal scaling behaviour in driven systems.

We study the coupled nonequilibrium dynamics of a conserved height field $h$ 
and a conserved signed density $\phi$ (that can be positive or 
negative, e.g., Ising spin-like degrees of freedom) within 
simple reduced models. In the absence of 
any general framework for nonequilibrium systems, such simple models are 
particularly useful 
 to study and answer questions of 
principle as we illustrate below. More 
specifically, we consider the nonlinearly coupled dynamics when $h$ is {\em 
autonomous}, i.e., the time-evolution of $h$ is independent of the second field 
$\phi$ and follows the C-KPZ equation. This models the dynamical evolution of a 
structureless signed species living on a fluctuating surface with conserved 
fluctuations. In the absence of the 
couplings with $h$, $\phi$ follows  
spatio-temporally scale invariant dynamics described by linear equations of 
motion with exactly known scaling exponents.  We consider two different 
models for conserved $\phi$-dynamics: (i) Model I, where the fluctuations of 
$h$ couples only with the 
spatial variation of $\phi$, given by
${\boldsymbol\nabla}\phi$, i.e., the dynamics of $\phi$ is invariant under 
the shift $\phi\rightarrow \phi 
+const.$, an {\em internal symmetry} that leaves the dynamics unchanged; 
and (ii) 
Model II, where the dynamics is {\em not} invariant 
under 
such a shift of $\phi$ (i.e., no such invariance, unlike in 
Model I). Generally, we find that the scaling 
properties  of Model I and Model II are starkly different - the spatio-temporal 
scaling of $\phi$ 
depends crucially on the detailed nature of its symmetry-determined coupling 
with $h$.
The 
remainder of the article is organised as follows: In Sec.~\ref{Model I-EOM}, 
we introduce Model I,
 write down the general symmetry permitted equations of motion 
for $h$ and $\phi$ and evaluate the scalings of the model parameters. Then in
Sec.~\ref{Model-II EOM}, we discuss Model II and note the 
differences between the two models. We finally
summarise and conclude in Sec.~\ref{Conclusion}.

\section{Model I}\label{Model I-EOM}

  The dynamics of 
$h$ is simply given by the C-KPZ equation~\cite{ckpz}
\begin{equation}
 \frac{\partial h}{\partial t}= -\nabla^2 \left[ \nu \nabla^2 h + 
\frac{\lambda_1}{2}(\nabla h)^2\right] + \eta_h,\label{ckpz}
\end{equation}
where $\eta_h$ is a Gaussian-distributed, zero-mean conserved noise with a 
variance $\langle\eta_h({\bf x},t) \eta_h(0,0)\rangle=-2 D_h \nabla^2 \delta ({\bf x}) 
\delta (t)$; $\nu>0$ is a damping coefficient and $\lambda_1$ is a nonlinear 
coupling constant~\cite{ckpz}.

We now write down the dynamical equations of $\phi$ in the 
hydrodynamic limit by using symmetry considerations. We
demand (i) translational and rotational invariance, (ii) conservation of 
$\phi$, and (iii) invariance under 
$\phi\rightarrow \phi +const$ for the dynamics of $\phi$. The last 
condition can be fulfilled only if derivatives of $\phi$ appear in the 
dynamical equations. Furthermore, for simplicity we restrict 
ourselves to systems that are linear in $\phi$-fluctuations, so that the 
dynamics of $\phi$ is invariant under the inversion of $\phi$. The 
general form of the relaxational equation of motion for a conserved density 
$\phi$ is (we ignore any advective processes)
\begin{equation}
 \frac{\partial\phi}{\partial t}= \mu\nabla^2 \frac{\delta {\mathcal 
F}}{\delta \phi} +{\rm NL}+\eta_\phi.\label{eqphi-basic}
\end{equation}
Here, ${\mathcal F}$ is a free energy functional that controls the dynamics and 
thermodynamics of $\phi$ in equilibrium. We choose 
\begin{equation}
{\mathcal F}=\int 
d^dx[ r_0\phi^2 + ({\boldsymbol\nabla}\phi)^2]/2,\label{free-en1}
\end{equation}
where we have neglected any 
nonlinear terms for simplicity; $r_0=T-T_c$ with $T$ as the temperature and 
$T_c$ the critical temperature. Furthermore, NL represents conserved 
nonlinear terms of 
nonequilibrium origin that are invariant under inversion of $\phi$ as well as 
a constant shift of $\phi$. We first consider the case with $r_0=0$, i.e., 
$\phi$-fluctuations are critical.
\begin{equation}
 \frac{\partial \phi}{\partial t}= 
 -\nabla^2 \left[ \mu \nabla^2 \phi + \lambda_2 ({\boldsymbol\nabla} 
h).({\boldsymbol\nabla} \phi) \right] + \eta_{\phi}.\label{eqphi1}
\end{equation}
 Here, $\eta_\phi$ is a Gaussian-distributed, zero-mean conserved noise with a 
variance
  $\langle\eta_{\phi}({\bf x},t) \eta_{\phi}(0,0)\rangle=-2 D_\phi \nabla^2 
\delta 
({\bf x}) \delta (t)$, $\mu>0$ is a damping coefficient and $\lambda_2$ is a 
non-linear cross-coupling coefficient through which $h$ affects the 
dynamics of $\phi$. The sign of $\lambda_2$ is arbitrary. 
Equation~(\ref{eqphi1}) corresponds to a  
current of $\phi$ given by
\begin{equation}
 {\bf J}_{\phi 1}= {\boldsymbol\nabla}[\mu\nabla^2\phi + \lambda_2 
({\boldsymbol\nabla h})\cdot({\boldsymbol\nabla}\phi)].\label{curr1}
\end{equation}
Thus, the nonequilibrium contribution to ${\bf J}_{\phi 1}$ can act only when 
both $\phi$ and $h$ have nonzero gradients and one of these gradients is 
spatially varying. 
Note that ${\bf J}_{\phi 1}$ remains invariant under 
$\phi \rightarrow \phi+const.$. This is in contrast to the models studied in 
Refs.~\cite{dro1,dro2}.
Furthermore, in contrast to our model I, the dynamics of the density field in
Ref.~\cite{dro2} is non-conserved.
Clearly, Eq.~(\ref{eqphi1}) is invariant under $\phi 
\rightarrow \phi + const.$. It is clear from the linearised versions of 
Eqs.~(\ref{ckpz}) and (\ref{eqphi1}) that the na\"ive scaling dimensions of $h$ 
and $\phi$ are identical.  Notice that both Eqs.~(\ref{ckpz}) and 
(\ref{eqphi1})
are invariant under spatial inversion, e.g., ${\bf x}\rightarrow -{\bf x}$, as 
well as inversion of $\phi$. 
Equations~(\ref{ckpz}) and (\ref{eqphi1}) do not admit any generalised Galilean 
invariance; see discussions below and Ref.~\cite{janssen} for technical 
comments.

\subsection{Scaling in Model I}

It is instructive to first consider the linearised version of 
Eq.~(\ref{eqphi1}) by 
setting 
$\lambda_2=0$. In that limit, the dynamics of $\phi$ can be solved {\em 
exactly}. In particular, in the Fourier space the correlation function $C_\phi 
({\bf q},\omega)=\langle |\phi({\bf q},\omega)|^2\rangle$ takes the form 
\begin{equation}
 C_\phi({\bf q},\omega)=\frac{2D_\phi q^2}{\omega^2 +\mu^2 q^8},\label{cphi1}
\end{equation}
where $\bf q$ and $\omega$ are the Fourier wavevector and frequency, 
respectively. Now, correlator (\ref{cphi1}) corresponds to the dynamic exponent 
$z_\phi =4$ and roughness exponent $\chi_\phi=\frac{2-d}{2}$ for the field 
$\phi$~\cite{stanley}. Compare these results with the correlations of 
$h$ from the linearised version of Eq.~(\ref{ckpz}). 
This yields the corresponding dynamic and roughness exponents for $h$ as 
$z_h=4$ and $\chi_h=\frac{2-d}{2}$, respectively. Clearly, $z_\phi=z_h$ at the linear level, 
implying 
{\em strong dynamic scaling} at the linear level. It is of course well-known 
that the scaling are affected by relevant (in a scaling sense) 
nonlinearities~\cite{halpin}, and as a result their values at the linear level 
get modified 
by the nonlinear effects. For instance, in the lowest order renormalised 
perturbation theory~\cite{ckpz}, $z_h=(12-\epsilon)/3$ with $\epsilon=2-d>0$, 
where as $z=4$ for $d\geq 2$~\cite{ckpz}. Whether or not strong 
dynamic 
scaling is still observed at the nonlinear level, is a question that we study 
here.

We can now write the dynamic generating functional~\cite{janssen1}, ${\mathcal Z}_I$, averaged 
over the 
noises 
$\eta_h$ and $\eta_\phi$, for the coupled system; see also 
Ref.~\cite{dro2} for similar functional approaches
\begin{equation}
 {\mathcal Z}_I=\int {\mathcal D} h {\mathcal D} \phi {\mathcal D} \hat h 
{\mathcal D} \hat \phi \exp[S_I], 
\end{equation}
where $\hat h$ and $\hat\phi$ are dynamic conjugate fields to $h$ and $\phi$, 
respectively~\cite{janssen1}; $S_I$ is the action functional given by

\begin{eqnarray}\label{action1}
 S_I&=& \int d^d x dt [ D_h \hat h \nabla^2 \hat h + D_\phi \hat\phi 
\nabla^2 \hat\phi \nonumber \\&+&
 \hat h \left( \frac{\partial h}{\partial t}+\nabla^2[\nu 
\nabla^2 h + 
\frac{\lambda_1}{2}(\nabla h)^2] \right)\nonumber \\
 &+& \hat \phi\left( \frac{\partial \phi}{\partial t}+\nabla^2[\mu \nabla^2 
\phi 
+ \lambda_2(\nabla h)(\nabla \phi)] \right)].
\end{eqnarray}

Nonlinear couplings $\lambda_1,\lambda_2$ preclude any exact enumeration of 
the relevant correlation functions from the action functional $S_I$ in
Eq.~(\ref{action1}). Naturally, perturbative calculations are used. Na\"ive 
perturbative expansions yield diverging corrections to the measurable 
quantities. In order to deal with these long wavelength divergences in a 
systematic manner,
we employ Wilson momentum shell dynamic renormalisation group
(DRG)~\cite{chaikin,halpin}. To this end, we first integrate out fields
$h({\bf q},\omega),\phi({\bf q},\omega)$ with wavevector
$\Lambda/b<q<\Lambda,\,b>1$, perturbatively up to the one-loop order in
(\ref{action1}). Here, $\Lambda$ is an upper cut off for wavevector. This allows
us to obtain the ``new" model parameters corresponding to a modified action
$S_I^<$
with an upper cutoff $\Lambda/b<\Lambda$; see Appendix for the corresponding 
one-loop Feynman diagrams.

In order to extract the renormalised parameters, we then rescale 
wavevectors
and frequencies according to ${\bf q}'=b{\bf q}$ and $\omega'=b^z{\bf \omega}$. 
Here $b=\exp[l]$ is a dimensionless length scale.
In a simple model with a single variable, $z$ becomes the dynamic 
exponent. For a multivariable problem as ours with the attendant possibility of 
unequal dynamic exponents for $h$ and $\phi$, the interpretation of $z$ in 
frequency rescaling as above will be clear as we go along. Under these 
rescalings, fields $h$ and $\phi$ also scale. We write, in Fourier space, $h({\bf q},\Omega)= \xi_h h({\bf q}^{\prime},\Omega^{\prime})$, 
$\hat h({\bf q},\Omega)= \hat \xi_h \hat h({\bf q}^{\prime},\Omega^{\prime})$, $\phi({\bf q},\Omega)= \xi_\phi \phi({\bf q}^{\prime},\Omega^{\prime})$,
$\hat \phi({\bf q},\Omega)= \hat \xi_\phi \hat \phi({\bf 
q}^{\prime},\Omega^{\prime})$. 
Using the redundancy~\cite{tauber} of the rescaling factors, $\hat \xi_h, \xi_h, 
\hat \xi_\phi$
and $\xi_\phi$, we impose the coefficients of  $\int d^d q d\Omega \hat h (-i\Omega) h$ 
and $\int d^d q d\Omega \hat \phi (-i\Omega) \phi$ to remain unity. 
This leads to the following condition on the rescaling factors :
\begin{equation}
 \hat \xi_h \xi_h=1=\hat \xi_\phi \xi_\phi.
\end{equation}

In the real space, let $h({\bf x}^{\prime}, t^{\prime})=\xi_h^R h(x,t)$ and 
$\phi({\bf x}^{\prime}, t^{\prime})=\xi_\phi^R \phi(x,t)$. Thus $\xi_h^R=b^{-(d+z)}\xi_h=b^{\chi_h}$ 
and $\xi_\phi^R=b^{-(d+z)}\xi_\phi=b^{\chi_\phi}$, where $\chi_h$ and $\chi_\phi$ are roughness exponents~\cite{tauber} associated with
$h$ and $\phi$, respectively. 

\subsubsection{Recursion relations and scaling exponents}

 We set up a perturbative DRG up to the one-loop order, where 
one-loop fluctuation corrections to the different model parameters are 
obtained. Notice that there are 
no fluctuation corrections to $\lambda_1$ at this order. In 
Ref.~\cite{ckpz}, 
this was ascribed to a modified Galilean invariance. Later on it was 
argued in Ref.~\cite{janssen} that there are indeed corrections to $\lambda_1$ 
at the two-loop order. Such considerations hold for Model I as well. Since we 
stick to a one-loop order DRG, we ignore such issues here.
 Following the standard DRG procedure~\cite{halpin,on}, we arrive at the 
following
recursion relations  [with $b=\exp [l]]$:

\begin{eqnarray}
 \frac{d \nu}{d l}&=& \nu[z-4+g (4-d)], \label{nufl} \\
 \frac{d \mu}{d l} &=& \mu[z-4+\frac{B^2 g}{P(1+P)}
 (4-d+\frac{2(1-P)}{1+P})],\label{mufl} \\
 \frac{d \lambda_1}{d l} &=& \lambda_1[z+\chi_h-4],\label{lam1fl} \\
 \frac{d D_h}{d l} &=& D_h[z-2-d-2\chi_h],\label{chihfl} \\
 \frac{d D_\phi}{d l} &=& D_\phi[z-2-d-2\chi_\phi],\label{chiphifl} \\
 \frac{d \lambda_2}{d l} &=& \lambda_2[\chi_h+z-4+ \frac{2gB(3+P)}{(1+P)^2}
 \nonumber \\ &-&\frac{4 g B^2}{(1+P)^2}-\frac{2gB}{1+P}], \label{lam3fl}
\end{eqnarray}
where $P=\frac{\mu}{\nu}, \, B=\frac{\lambda_2}{\lambda_1}$ and 
$g=\frac{\lambda_1^2 D_h K_d \Lambda^2}{4 \nu^3 d}$ are the
effective dimensionless coupling constants; under rescaling of space and time 
$g$ scales as $ b^{2-d}$ implying $d=2$ to be the critical 
dimension~\cite{ckpz}. The 
flow equations for $g,P$ and $B$ may be immediately obtained:

\begin{eqnarray}\label{eff_rec}
 \frac{d g}{d l} &=& g[2 -d + g (d-4)],
\nonumber \\
 \frac{d P}{d l}&=&-Pg[4-d - \frac{2 
B^2}{P(1+P)}(4-d+\frac{2(1-P)}{1+P})],\nonumber \\
 \frac{d B}{d l}&=&B[2gB\frac{(3+P)}{(1+P)^2}-\frac{4gB^2}{(1+P)^2}
 -\frac{2Bg}{1+P}].\label{couple-flow}
\end{eqnarray}

At the DRG fixed point (FP), $dg/dl=0=dB/dl=dP/dl$. Then, 
we have from Eq.~(\ref{eff_rec}), $B^*=1, P^*=1$ and $ g^*=\frac{2-d}{3(4-d)}$ 
or $g^*=0$ at 
the FP. Linear stability analysis reveals that $g^*= 
\frac{2-d}{3(4-d)}$ is the stable FP for $d< 2$; for $d\geq 2$, 
$g^*=0$~\cite{ckpz}. For $d< 2$ and with these values of $B^*, P^*$ and 
$g^*$ at the stable DRG FP, we note 
that both (\ref{nufl}) and (\ref{mufl}) yield the choice $z=\frac{10+d}{3}$ at 
the DRG FP make both $d\nu(l)/dl$ and $d\mu(l)/dl$ zero.  This in turn 
implies 
that both $h$ and $\phi$ have the same dynamic exponent 
$z_h=z_\phi=\frac{10+d}{3}$. 
Thus, Model I displays {\em strong dynamic scaling}. Furthermore, by using 
(\ref{chihfl}) and (\ref{chiphifl}) at the stable DRG FP for $d<2$, we obtain 
$\chi_h=\chi_\phi=\frac{2-d}{3},\,d< 2$. 
Also, expectedly in contrast to the results in Ref.~\cite{dro2},
the flat phase of the CKPZ equation becomes unstable below $d=2$ and not 
below $d=4$, due to the roughness exponent
becoming positive  below $d=2$~\cite{ckpz}; equivalently, $d=2$ is the 
critical dimension for the CKPZ equation.
For $d\geq 2$, the nonlinearties 
are irrelevant (in a RG sense) and hence the results from the linear theory 
holds.Note that the nonlinearities become irrelevant in Ref.~\cite{dro2}
only above $d=6$.

\subsubsection{Model I with $\lambda_1=0$} 

 Consider now the limiting case with $\lambda_1=0$.  
Thus, $h$ 
evolves linearly with $z_h=4$ and $\chi_h=\frac{2-d}{2}$ known exactly.  The 
flow 
equations simplify to

\begin{eqnarray}\label{flow-reduc}
 \frac{d \mu}{d l} &=& \mu[z_\phi-4+\frac{\lambda_2^2 D_h K_d}{2\nu\mu(\nu+\mu) 
d}(4-d +\frac{2(\nu-\mu)}{\nu+\mu})],\\ \nonumber
 \frac{d D_\phi}{d l} &=& D_\phi[z_\phi-2+d-2\chi_\phi], \\
 \frac{d \lambda_2}{d l} &=& \lambda_2[\chi_h+z_\phi-4-\frac{\lambda_2^2 D_h 
K_d \Lambda^2}{\nu(\mu+\nu)^2 d}]. \nonumber 
\end{eqnarray}
In obtaining the flow equations (\ref{flow-reduc}), we have rescaled time $t$ 
that corresponds to a dynamic exponent $z_\phi$.

 Clearly, there are positive corrections to $\mu$. Thus, 
scale-dependent $\mu(l)\gg \nu (l)=\nu$, as the DRG FP is approached. Thus, we 
already conclude that $z_\phi < z_h=4$. Hence, {\em weak dynamic scaling} is 
expected, implying $\nu (l)/\mu(l)\rightarrow 0$ as $l\rightarrow 
\infty$. In that limit we find from the above flow equations
\begin{eqnarray}
\frac{d \mu}{d l} &=& \mu[z-4+\frac{\lambda_2^2 D_h K_d}{2\nu\mu^2 d}(2-d)],\\
 \frac{d \lambda_2}{d l} &=& \lambda_2[\chi_h+z-4-\frac{\lambda_2^2 D_h K_d \Lambda^2}{\nu\mu^2 d}]
\end{eqnarray}
We identify an effective coupling constant 
$\tilde g=\frac{\lambda_2^2 D_h K_d \Lambda^2}{\nu\mu^2 d}$ that scales as 
$b^{2-d}$ under the rescaling of space and time. This shows that $d=2$ is the 
critical dimension, such that for $d<2$ fluctuation corrections should be 
relevant in the long wavelength limit. The DRG flow 
equation for $\tilde g$ is 
\begin{equation}
 \frac{d\tilde g}{dl}=\tilde g [2-d+2\tilde g(d-3)].
\end{equation}
At the DRG FP, $d\tilde g/dl=0$, yielding $\tilde g=\frac{2-d}{2(3-d)}$ as the 
stable FP for $d<2$, where as $\tilde g=0$ is the stable FP for $3>d>2$. The 
apparent singularity in $\tilde g$ at $d=3$ is likely to be an artifact of a 
low order perturbation theory used here~\cite{kpz-singu}.   
This then implies 
$z_\phi=4+\frac{(2-d)^2}{2d-6}=4+O(\epsilon)^2,\chi_\phi=\frac{3d(d-2)-8}{4(d-3)
} $. 
Thus, $z_\phi$ differs from $z_h$ by $O(\epsilon)^2$. Since our one-loop 
analysis is valid only up to $O(\epsilon)$, we set $z_\phi=z_h$ at this order, 
restoring strong dynamic scaling. Whether or not this remains true at higher 
order remains to be checked. On the whole, thus, within a one-loop 
approximation Model I displays strong 
dynamic scaling independent of whether nonlinear effects are considered in the 
dynamics of $h$, i.e., $\lambda_1=0$ or not. Whether or not this remains true 
at higher order remains to be seen. In contrast, Ref.~\cite{dro1} finds 
both equal ($z_h=z_\phi$) and unequal   ($z_h\neq z_\phi$) dynamic exponents 
(at $d=1$), 
depending upon the details of the nonlinear couplings. Furthermore, 
Model I has $d=2$ as the critical dimension, similar to the KPZ 
equation~\cite{natter}, or the conserved KPZ equation~\cite{ckpz}. In contrast, 
the interplay between the KPZ surface fluctuations and phase separation dynamics 
tend to make the critical dimension higher, as reported in Ref.~\cite{dro2}. 
Unsurprisingly, the scaling behaviour of Model I is completely different from 
those in Ref.~\cite{dro2}.

\subsubsection{Model I with $r_0>0$}

We now briefly discuss the dynamics of $\phi$ for $r_0>0$, i.e., 
$\phi$-fluctuations are noncritical. This generates a linear $\nabla^2\phi$ 
term in Eq.~(\ref{eqphi-basic}) leading to
\begin{equation}
  \frac{\partial \phi}{\partial t}= 
 -\nabla^2 \left[-\mu_1\phi+ \mu \nabla^2 \phi + \lambda_2 ({\boldsymbol\nabla} 
h).({\boldsymbol\nabla} \phi) \right] + \eta_{\phi},\label{eqphi1-2}
\end{equation}
where $\mu_1=\mu r_0>0$. Equation (\ref{eqphi1-2}) gives $z_\phi=2$ at the 
linear level, corresponding to weak dynamic scaling, since the 
$\mu_1\nabla^2\phi$-term is more relevant than the $-\mu\nabla^4\phi$-term (in 
a scaling sense).  However, with the existing form of 
the $\lambda_2$-nonlinear term, corrections to the propagator are still all at 
$O(q^4)$.  This implies that there are no fluctuation-corrections to the 
$\mu_1$-term, yielding $z_\phi=2$ {\em exactly} even at the nonlinear 
level, and hence weak dynamic scaling prevails. This together with the exact 
knowledge 
of $\chi_\phi$ from the non-renormalisation of $D_\phi$ yield the scaling 
exponents of $\phi$ exactly, which are identical to their values in the 
corresponding linear theory. Dynamics of $\phi$, then, is totally unaffected by 
the nonequilibrium drive when $\phi$-fluctuations are noncritical.

 \section{Model II} \label{Model-II EOM}
 
  In Model I above, height fluctuations couple with 
${\boldsymbol\nabla}\phi$, the local spatial variation in $\phi$.  In contrast,
 we now consider the case when ${\boldsymbol\nabla}h$ couples directly 
with $\phi$; consequently the dynamics is {\em not invariant} under a constant 
shift of $\phi$: $\phi\rightarrow \phi + const.$.  We again consider the case 
where the dynamics of 
$h$ is autonomous, i.e., unaffected by $\phi$. Thus, the dynamical equation of 
$h$ is still given by Eq.~(\ref{ckpz}). The dynamical equation for $\phi$ is 
still given by Eq.~(\ref{eqphi-basic}), while the nonequilibrium terms NL 
should now include conserved terms that break the symmetry under a constant shift 
of $\phi$ as well. We continue to assume that the dynamics in linear in $\phi$. 
Furthermore, we now set $r_0>0$, i.e., $T>T_c$ (hence 
$\phi$ is 
noncritical); we briefly discuss the $r_0$ case at the end. With 
all these, 
the most general equation for $\phi$ to 
the leading order in nonlinearities and spatial gradients in the hydrodynamic 
limit is now given by
 \begin{equation}\label{phi2_eq}
  \frac{\partial \phi}{\partial t} = \tilde \mu \nabla^2 \phi + g_2 
{\boldsymbol\nabla} 
(\phi {\boldsymbol\nabla} h)+\eta_\phi.
 \end{equation}
  Here, $\tilde\mu>0$ is a damping coefficient, $g_2$ is a nonlinear coupling 
constant; noise $\eta_\phi$ is same as that in Model I. Notice that the 
nonlinear coupling term $g_2$ is identical to the one introduced in 
Ref.~\cite{dro1}.
Equation~(\ref{phi2_eq}) corresponds to a current
\begin{equation}
 {\bf J}_{\phi 2}=-\tilde\mu{\boldsymbol\nabla}\phi - g_2  
\phi {\boldsymbol\nabla} h.\label{curr2}
\end{equation}
Thus, the nonequilibrium parts in ${\bf J}_{\phi 2}$ contribute wherever there 
is a local tilt in the surface given by ${\boldsymbol\nabla}h$ with a local 
$\phi$~\cite{dro1}. This distinguishes the 
nonequilibrium effects of Model II from Model I.
Both Eqs.~(\ref{phi2_eq}) and (\ref{curr2}) are
clearly  {\em not} invariant under $\phi\rightarrow\phi 
+const.$. At this stage it is 
convenient to split $\phi$ as a sum of its mean $\phi_0=\int d^dx \phi({\bf 
x},t)/V$ and a zero-mean fluctuating part; here $V$ is the system volume. This
clearly generates a linear term proportional to $\phi_0\nabla^2 h$.   Such a 
term 
manifestly breaks the symmetry under inversion of $\phi$. In effect, 
$\phi_0$ now parametrises the dynamics of $\phi$. We set $\phi_0=0$ and denote 
the fluctuating part with zero-mean by $\phi$ below. This then
restores the symmetry under inversion of $\phi$.
 Notice that 
Eq.~(\ref{phi2_eq}) is invariant under ${\bf x}\rightarrow -{\bf x}$.

\subsection{Scaling in Model II}

 Similar to our analysis for Model I, we first consider the 
 linearised version of (\ref{phi2_eq}) 
that can be solved exactly. 
 The correlation function $C_{\phi}({\bf 
q}, \omega)$ then takes the exact form:
  \begin{equation}
  C_\phi({\bf q},\omega)=\frac{2D_\phi q^2}{\omega^2 +\tilde\mu^2 
q^4}.\label{cphi2}
 \end{equation}
 Equation~(\ref{cphi2}) implies that $z_\phi=2$ and roughness exponent 
$\chi_\phi=-\frac{d}{2}$ for the density field $\phi$. Thus, at the linear level,
 this clearly implies {\em weak} dynamic scaling since $z_h =
4\neq z_\phi$ at the linear level. As before, we study whether and
 if so, how the nonlinear effects modify these scaling behaviors, and in 
particular if weak dynamic scaling can get further reinforced (larger 
differences between $z_h$ and $z_\phi$) or otherwise by nonlinear effects.
 \par
 
 The action functional $S_{II}$ for Model II is given by
 
 \begin{eqnarray}\label{action2}
 S_{II}&=& \int d^d r dt [ D_h \hat h \nabla^2 \hat h + D_\phi \hat\phi 
\nabla^2 \hat\phi \nonumber \\&+&
 \hat h \left( \frac{\partial h}{\partial t}+\nabla^2[\nu \nabla^2 h + 
\frac{\lambda_1}{2}(\nabla h)^2] \right)\nonumber \\
 &+& \hat \phi\left( \frac{\partial \phi}{\partial t}-\tilde\mu \nabla^2 \phi
- g_2\nabla(\phi \nabla h) \right) ].
 \end{eqnarray}
  As in Model I, nonlinearities preclude any exact enumeration of 
the scaling exponents. We again resort to perturbative DRG up to the one-loop 
order as for Model I.
 
\subsubsection{Rescaling of fields and parameters: recursion relations and 
scaling exponents}
 
 We enumerate the one-loop corrections to the various model parameters in Model 
II. See Appendix for the one-loop diagrams for Model II. We rescale time by a 
factor that corresponds to a dynamic exponent $z$. The interpretation of $z$ as 
the dynamic exponent of $h$ or $\phi$ will become clear below.
The recursion relations
for the model parameters are given by: 
 
 \begin{eqnarray}
  \frac{d \nu}{d l}&=& \nu[z-4+g(4-d)], \nonumber \\
  \frac{d \tilde \mu}{d l} &=& \tilde \mu [z-2+g_3],\nonumber \\
  \frac{d \lambda_1}{d l} &=& \lambda_1[z+\chi_h-4],\nonumber \\
 \frac{d D_h}{d l} &=& D_h[-d+z-2-2\chi_h],\nonumber \\
 \frac{d D_\phi}{d l} &=& D_\phi[-d+z-2-2\chi_\phi+g_3],\nonumber \\
 \frac{d g_2}{d l} &=& g_2[z-2+\chi_h-g_3],\label{flow-model2}
 \end{eqnarray}
 where $g=\frac{\lambda_1^2 D_h K_d \Lambda^2}{4 \nu^3 d}$ (same as in Model I) 
and 
 $g_3=\frac{g_2^2 D_h K_d \Lambda^2}{\nu \tilde \mu^2 d}>0$ are the 
 effective dimensionless coupling constants for Model II that scales as 
$b^{2-d}$ under rescaling of space and time, suggesting $d=2$ to be the 
critical dimension (same as in Model I).
 Now, if we set $\frac{d \tilde \mu}{d l}=0$ at the DRG FP, we find
 \begin{equation}\label{z,C}
  z= 2-g_3.
 \end{equation}
 Since this choice of $z$ leaves $\tilde \mu(l)$ scale independent, we identify it as 
the dynamic exponent of $\phi$: $z_\phi=z$. Notice that this choice for $z$ 
{\em does not} leave $\nu(l)$ scale independent, and hence cannot be the 
dynamic exponent $z_h$ of $h$ (which is anyway known independently, 
$z_h=(10+d)/3$ 
for $d\leq 2$ and $z_h=4$ for $d\geq 2$). With this choice for $z$, instead 
$\nu(l)$ scales as $l^{z-4+g(4-d)}$~\cite{weak}. This scale-dependent $\nu(l)$ 
together with the identification $l=-\ln q$ may be used to extract $z_h$ from 
the formal definition 
\begin{equation}
 C_h(q,\omega)=\frac{2D_h q^2}{\omega^2 + \nu(q)^2 q^{2z}},
\end{equation}
see Ref.~\cite{weak} for more details. This line of argument yields the same 
value for $z_h$, as already obtained above.

 Equation~(\ref{z,C}) combined with $\frac{d D_\phi}{d l}=0$, gives 
$\chi_\phi=-\frac{d}{2}$ at the DRG FP.
 Note that $\chi_h$ and $z_h$ retain respectively the same values as in Model I.
  \par
  Further, the 
DRG flow equation for $g_3$ is
    \begin{equation}
   \frac{d g_3}{ d l}= 2 g_3 (\chi_h-2 g_3).
  \end{equation}
 Thus, we have at the FP, $g_3=0,\frac{\chi_h}{2}$. Since $\chi_h=(2-d)/3$, 
$g_3=(2-d)/6$ gives the stable FP for $d<2$; for $d\geq 2$, $g_3=0$ 
at the stable FP. Thus for $d\geq 2$, the coupling becomes irrelevant in the 
dynamics of
  $\phi$ and the results from the linear equation holds. But for $d<2$, 
nonlinear effects are relevant and 
$z_\phi=2-g_3=2-\frac{\chi_h}{2}=\frac{10+d}{6}$. In general, for any $d$ weak 
dynamic 
scaling prevails in the system.

 \subsubsection{Dynamics of $\phi$ when $\lambda_1=0$}
 
 We now consider the case when $\lambda_1=0$ for model II. In this limit, $\frac{d\nu}{dl}=0 \implies z_h=4$.
 From the flow equation of $g_2$ and $g_3$, we have
 
 \begin{equation}
  \frac{d g_2}{dl}=g_2[z-2+\chi_h -g_3]
 \end{equation}
 and
 
 \begin{equation}
  \frac{d g_3}{dl}=2 g_3 (\chi_h - 2 g_3).
 \end{equation}

 Now in the limit of $\lambda_1=0$, we have
 
 \begin{equation}\label{chi-limit}
  \chi_h=\frac{2-d}{2}.
 \end{equation}

 Using the fact that at FP, either $g_3=0$ (stable for $d\geq 2$), or 
$g_3=\frac{\chi_h}{2}$ (stable for $d<2$) and the value of $\chi_h$ given by
 Eq.~\ref{chi-limit}, we arrive at two values for $z_\phi$.
 \par
 For $g_3=0, z_\phi =2$ and $\chi_\phi=-d/2$, valid for $d\geq 2$. When 
$g_3=\frac{\chi_h}{2}, z_\phi = \frac{6+d}{4}$ and $\chi_\phi=-d/2$, valid for 
$d<2$. Thus $\chi_\phi=-d/2$ for all $d$.
 
The results from Model II are complementary to those in 
Ref.~\cite{dro1}. While Ref.~\cite{dro1} considered
non-conserved dynamics for $h$ and conserved dynamics for $\phi$ at $d=1$, we 
have 
considered conserved dynamics for both the
fields $h$ and $\phi$ in Model II in general $d$ dimensions, with the dynamics 
of $h$ being treated as autonomous for simplicity. Nonetheless, Model 
II and the studies in Ref.~\cite{dro1} display universal scaling very different 
from each other - a hallmark of the models being nonequilibrium and unlike 
equilibrium models where a conservation law can affect only the dynamic scaling 
behaviour; see, e.g., model A and model B in the language of Ref.~\cite{halpin}. 
In contrast, none of the scaling exponents in the two studies have any simple 
relations. While 
the roughness exponent takes the value
$\frac{-d}{2}$ for all $d$ in our Model II, it can take several values depending upon the model parameters in Ref.~\cite{dro1}.
More importantly, Ref.~\cite{dro1} shows the possibility of both strong and 
weak dynamic scalings. In sharp contrast, our Model II
only gives weak dynamic scaling for any $d$ and for both the stable and unstable FP values of the coulping constant, $g_3$. Thus
comparison between the results of Ref.~\cite{dro1} and Model II significantly establishes how conservation laws can lead
to entirely different physical outcomes, even though the coupling 
between the degrees of freedom can have the same structure.

\subsubsection{Model II with $r_0=0$}

We briefly discuss what happens when $r_0=0$, i.e., the $\phi$-fluctuations are 
critical. With $r_0=0$, the $\tilde\mu\nabla^2\phi$-term in Eq.~(\ref{phi2_eq}) 
is to be replaced by a $\nabla^4\phi$-term. Nonetheless, with the existing 
nonlinear term in (\ref{phi2_eq}), the lowest order corrections to the 
propagator are at $O(q^2)$, thus generating a $\nabla^2\phi$-term in the 
fluctuation corrected equation. All our results for Model II with $r_0>0$ derived above then 
immediately follow. It is however possible to start with a specific {\em bare} 
$r_0$, so that the fluctuation-corrected $r_0$ vanishes. The relaxation of 
$\phi$-fluctuations will now be controlled by a (subleading to a bare 
$\tilde\mu\nabla^2\phi$-term) $\nabla^4\phi$-term, with a $z_\phi=4$ at the 
linear level (as in Model I with $r_0=0$). However, the fluctuation corrections 
to  this $\nabla^4\phi$-term are expected to be different from those in Model I 
with $r_0=0$, owing to the different form of the nonlinear term in 
Eq.~(\ref{phi2_eq}). We do not discuss the details here.
 
 \section{Conclusions and outlook} \label{Conclusion}

 We have thus investigated how the presence or absence of an internal 
symmetry affects the universal scaling properties in the noisy dynamics of a 
conserved scalar 
density  driven by a fluctuating conserved KPZ surface $h$. We make a 
particularly simple choice for internal symmetry, {\em viz.} invariance under a 
constant shift of $\phi$. To this end, we consider 
two specific reduced models, Model I and Model II, to address how the interplay 
between the symmetries that control the structure of the nonlinear terms
and conservation laws control the universal scaling properties. In Model I, 
$h$-fluctuations couples with 
${\boldsymbol\nabla}\phi$, rendering the ensuing dynamics of $\phi$ 
independent  of $\phi_0$, the mean of $\phi$. Model I is constructed in way to 
respect the invariance under inversion of $\phi$.
At the linearised level, both $h$ and $\phi$ dynamics display strong dynamic 
scaling with a single dynamic exponent $z=4$. Beyond the linearised theory, the 
 scaling exponents depend crucially on the details of the nonlinear 
couplings, and also whether $\phi$ is a critical field or noncritical.
 The relevant scaling exponents are evaluated in a one-loop DRG 
calculation for critical $\phi$-fluctuations; for noncritical 
$\phi$-fluctuations, the scaling exponents are unaffected by the nonlinearity 
and known exactly, that corresponds to weak dynamic scaling. For critical 
$\phi$-fluctuations, strong dynamic scaling ensues.

We have studied another model, Model II, where $h$-fluctuations directly 
couples with $\phi$. Thus in contrast to Model I, Model II does not remain invariant under a
constant shift of $\phi$. As a result, $\phi_0$, the mean of $\phi$, parametrises 
the dynamics of $\phi$. 
We focus on the particular case where $\phi_0=0$. This restores the symmetry of 
the model under inversion of $\phi$-fluctuations. In Model II (with $r_0>0$), 
even at the 
linear level weak dynamic scaling follows ($z_\phi=2$), a feature that holds 
good even when 
the nonlinear effects are taken into account.  
Furthermore, if we assume $\phi_0\neq 0$, we obtain additional linear term 
proportional to $\nabla^2 h$ in (\ref{phi2_eq}). Given that the dynamics of $h$ 
is independently known (being autonomous), this term effectively acts like an 
additional {\em additive noise} in the problem, whose correlation is {\em not} 
$\delta$-correlated in space and time. This is likely to affect the scaling 
properties of $\phi$ in nontrivial ways. Comparison of the results from Model I 
and Model II thus establish the significance of the internal symmetry under 
constant shifts of $\phi$ in determining the scaling properties.
It would be of interest to construct equivalent discrete lattice-gas models and 
study these issues there. The specific microscopic rules for the lattice-gas 
models for Model I and Model II may be formed from the nonequilibrium 
contributions to the currents (\ref{curr1}) and (\ref{curr2}) respectively. We 
welcome further work along this direction.

We now make a brief general comparison of our studies here with those on 
the generalised 
coupled KPZ equations by DK. First and foremost, DK allowed for the feedback 
of the density fluctuations on the height fluctuations, whereas in our 
case, the the dynamics of the surface is assumed to be autonomous, 
independent of the density fluctuations. Furthermore, the surface fluctuations 
in both Model I and Model II in our studies are conserved, and hence slower 
than the nonconserved height fluctuations in the models of DK. This is reflected 
in the generic higher values of the dynamic exponents $z_h$ in our studies.

For reasons of simplicity, we have ignored a $\phi^4$ term in 
the free energy (\ref{free-en1})~\cite{comment1} while setting up Model I or 
Model II. Such a term, if included, will generate a $\sim \phi^3$ term in the 
dynamics of $\phi$ in both Model I and Model II.  Clearly, this term 
manifestly breaks
the $\phi \rightarrow \phi + const.$ symmetry of Model I. Thus all the 
couplings 
present in
Model II that also break the invariance under a constant shift of $\phi$ should 
now be included, and Model I will effectively reduce to Model II (albeit at 
$r_0=0$). In case of Model II, a $\phi^3$-term in the dynamics of $\phi$  
will lead to competition with  the already existing
nonlinearities in Model II; the resulting scaling behaviour can further be 
investigated within the framework
of one-loop RG (not done here).

Our consideration of the the dynamics of $h$ as autonomous 
is clearly a limiting case. More generally, generic nonlinear feedback of 
$\phi$ on 
the dynamics of $h$ may be present. Again due to symmetry reasons the nonlinear 
structure of the 
feedback 
should differ from Model I to Model II. It would be interesting to see 
whether and how the feedback may alter the conclusions drawn above. Furthermore, our 
equations of motion are all invariant under spatial inversion. An interesting 
generalisation
would be to allow terms that violate this invariance under spatial inversion. Such terms
may potentially lead to generation of underdamped kinematic waves, absent in Model I
or Model II. Kinematic waves can be important, e.g., these waves lead to
weak dynamic scaling in Ref.~\cite{weak}. Whether similar breakdown 
of strong dynamic 
scaling occurs in Model I in the presence of kinematic waves remains 
to be investigated. The dynamical field $\phi$ being a conserved density 
follows a conservation law form of equation of motion. Similar to 
Ref.~\cite{dro2}, phase ordering dynamics on a conserved KPZ surface may be 
studied by making $\phi$ a nonconserved density.  A nontrivial variant of 
this would be to consider $\phi$ to be a broken symmetry mode that follows a 
nonconserved equation of motion, but executes a scale-invariant dynamics. When 
such a broken symmetry variable is driven by a conserved KPZ field, the 
emerging scaling properties are likely to be quite different from what is 
reported here. Lastly, coupled systems with linear instabilties may be 
considered, such that the nonequilibrium steady states may even involve 
patterns. We look forward to future work in these directions.

\section{Acknowledgement}

The authors gratefully acknowledge the Alexander Von Humboldt Stiftung 
for partial financial support through the Research Group Linkage Programme 
(2016).

\appendix
\section{ Model I: Results}
 In the Fourier space, the propagators and correlators of $h$ and $\phi$, 
respectively, thus take the following form :
 
 \begin{eqnarray}
  \langle\hat h({\bf q},\omega)h(-{\bf q},-\omega)\rangle&=&\frac{-1}{-i\omega 
+\nu q^4}\nonumber \\
  \langle h({\bf q},\omega)h(-{\bf q},-\omega)\rangle&=& \frac{2 D_h 
q^2}{\omega^2 + \nu^2 q^8}\nonumber 
\\
  \langle\hat \phi({\bf q},\omega)\phi(-{\bf 
q},-\omega)\rangle&=&\frac{-1}{-i\omega +\mu q^4}\nonumber 
\\
  \langle\phi({\bf q},\omega)\phi(-{\bf q},-\omega)>&=& \frac{2 D_\phi 
q^2}{\omega^2 + \mu^2 
q^8}. 
 \end{eqnarray}

\subsection{One loop corrections to the model parameters}

The corrections of the model parameters and the corresponding relevant Feynman diagrams for Model I are given below :

\begin{figure}[htb]
\includegraphics[width=7cm]{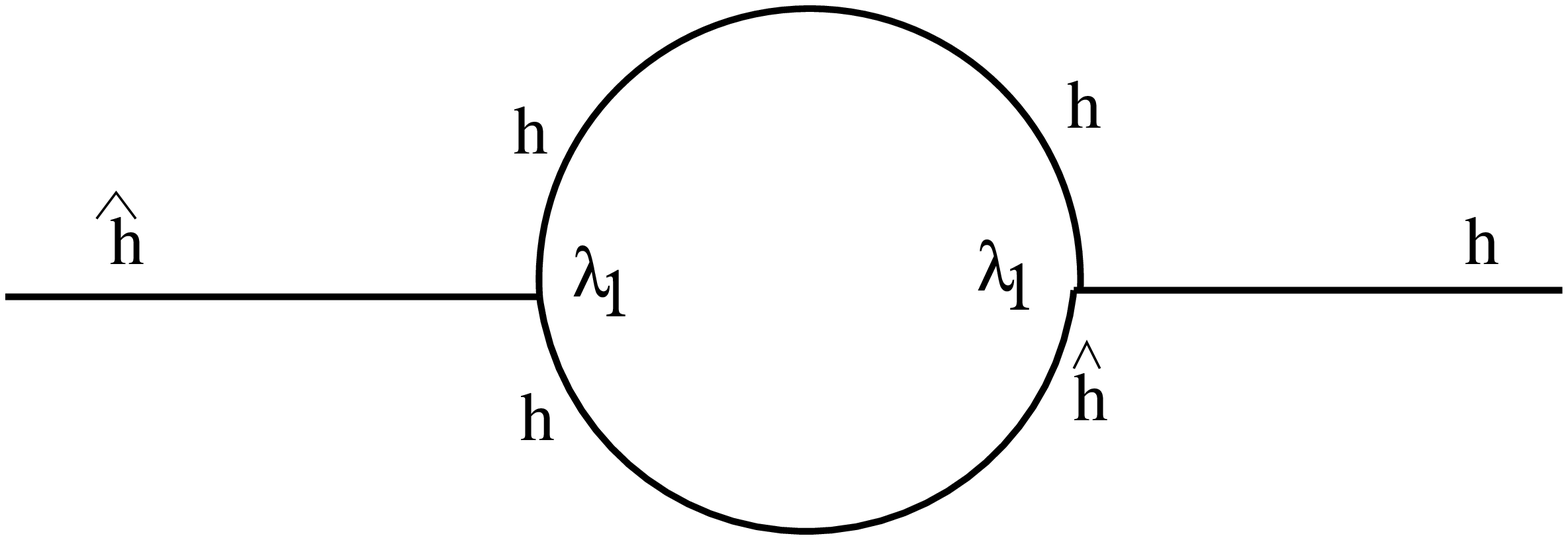}\hfill
\caption{One loop correction to $\nu$.}  \label{nu-cor}
\end{figure}

\begin{equation}\label{nu_corr}
 \nu^{<}=\nu -\frac{\lambda_1^2 D_h K_d}{\nu^2}\left[ 
\int_{\Lambda/b}^{\Lambda}\frac{d q \, q^{d-1}}{4 q^2} -
 \int_{\Lambda/b}^{\Lambda}\frac{d q \, q^{d-1}}{ q^2 d} \right]
\end{equation}

Fig.~\ref{nu-cor} shows the relevant Feynman diagram for one-loop correction to $\nu$.

\begin{figure}[htb]
\includegraphics[width=7cm]{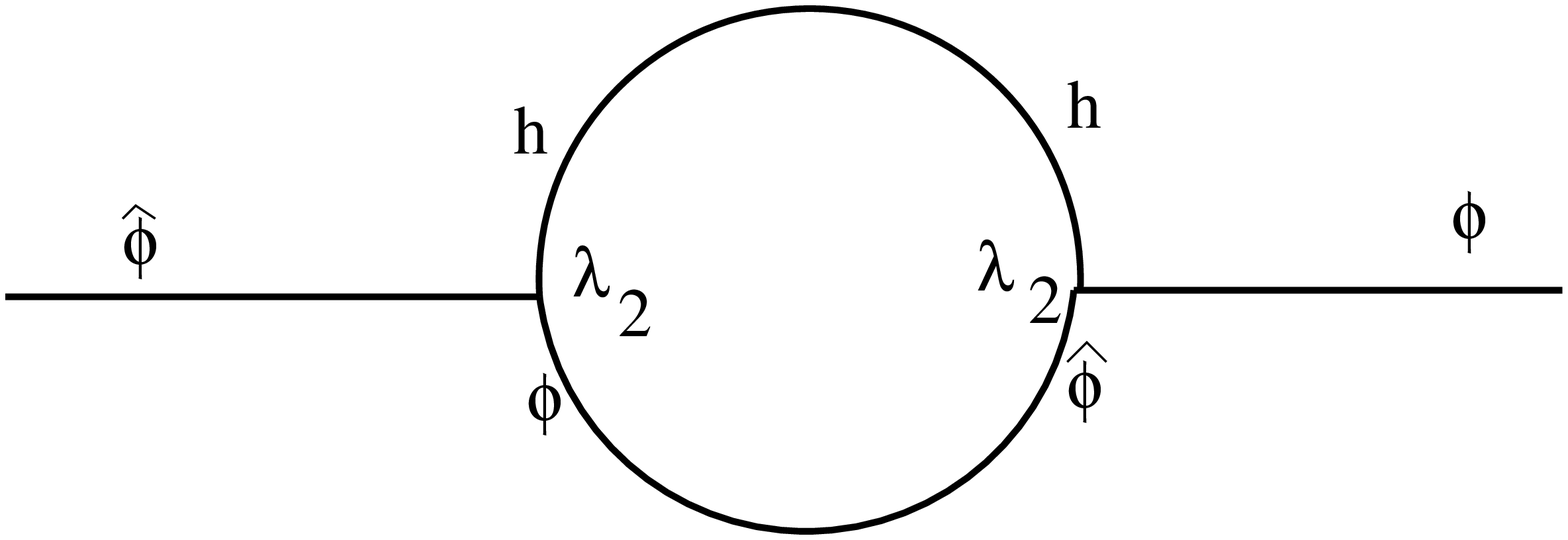}\hfill
\caption{One loop correction to $\mu$.}  \label{mu-cor}
\end{figure}

\begin{equation}\label{mu_corr}
 \mu^{<}=\mu -\frac{\lambda_2^2 D_h K_d}{\nu(\nu+\mu)}\left[ 
\frac{1}{2}-\frac{2+(\nu-\mu)/(\nu+\mu)}{d}\right]
 \int_{\Lambda/b}^{\Lambda}\frac{d q \, q^{d-1}}{q^2} 
\end{equation}

See Fig.~\ref{mu-cor} for the Feynman diagram corresponding to one-loop correction to $\mu$.

\begin{equation}\label{lam1_corr}
\lambda_1^{<}=\lambda_1
\end{equation}

\begin{equation}\label{Dh_corr}
 D_h^{<}=D_h
\end{equation}

\begin{equation}\label{Dphi_corr}
 D_{\phi}^{<}=D_{\phi}
\end{equation}

\begin{figure}[htb]
\includegraphics[width=8cm]{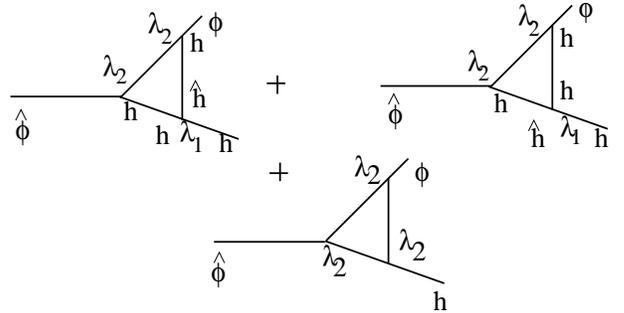}\hfill
\caption{One loop correction to $\lambda_2$.}  \label{lam2-cor}
\end{figure}

\begin{eqnarray}\label{lam3_corr}
 \lambda_2^{<}&=&\lambda_2 +  [ 
 \frac{D_h \lambda_2^2 \lambda_1 K_d(3 \nu +\mu)}{2 \nu^2 (\nu +\mu)^2 d}
- \frac{D_h \lambda_2^3 K_d}{\nu (\nu +\mu)^2 d}
\nonumber \\ &-&\frac{D_h \lambda_2^2 \lambda_1 K_d}{2 \nu^2 (\nu +\mu) d}]
\int_{\Lambda/b}^{\Lambda}\frac{d q \, q^{d-1}}{q^2}
\end{eqnarray}

Relevant one-loop corrections to $\lambda_2$ are given in Fig.~\ref{lam2-cor}.

\subsection{Model II results}

The propagators and correlators for the system are given by

 \begin{eqnarray}
  \langle\hat h({\bf q},\omega)h(-{\bf q},-\omega)\rangle&=&\frac{-1}{-i\omega 
+\nu q^4}\nonumber \\
  \langle h({\bf q},\omega)h(-{\bf q},-\omega)\rangle&=& \frac{2 D_h 
q^2}{\omega^2 + \nu^2 q^8}\nonumber \\
  \langle \hat \phi({\bf q},\omega)\phi(-{\bf 
q},-\omega)\rangle&=&\frac{-1}{-i\omega + \tilde \mu q^2}\nonumber \\
  \langle\phi({\bf q},\omega)\phi(-{\bf q},-\omega)\rangle&=& \frac{2 D_\phi 
q^2}{\omega^2 + \tilde \mu^2 q^4}.  
 \end{eqnarray}

 As before, we now find corrections to the bare model parameters by evaluating integrals upto one-loop order
 from wavevector $\Lambda/b$ to $\Lambda$. This leads to the following
 results :
 
 \begin{equation}\label{nu2_cor}
  \nu^{<}=\nu -\frac{\lambda_1^2 D_h K_d}{\nu^2}\left[ \int_{\Lambda/b}^{\Lambda}\frac{d q \, q^{d-1}}{4 q^2} -
 \int_{\Lambda/b}^{\Lambda}\frac{d q \, q^{d-1}}{ q^2 d} \right]
 \end{equation}

 \begin{equation}\label{til-mu_cor}
  \tilde \mu^{<}=\tilde \mu + \frac{g_2^2 D_h K_d}{\nu \tilde \mu d} \int_{\Lambda/b}^{\Lambda}\frac{d q \, q^{d-1}}{q^2}
 \end{equation}

 \begin{figure}[htb]
\includegraphics[width=8cm]{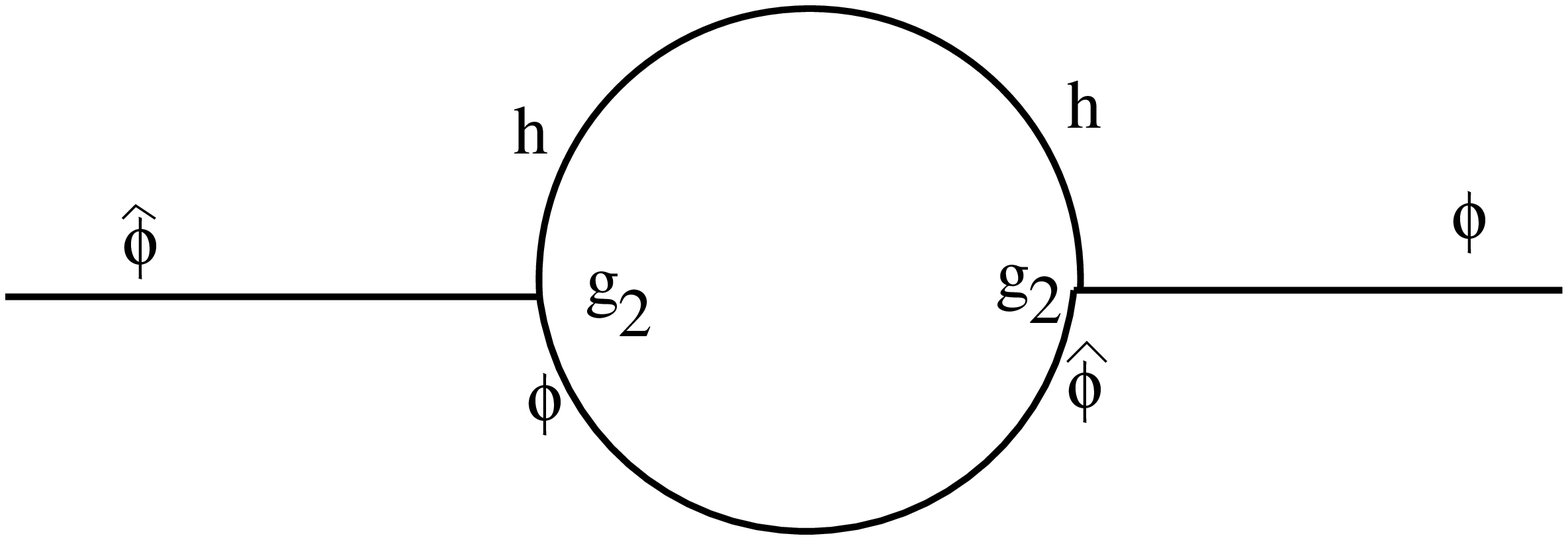}\hfill
\caption{One loop correction to $\tilde \mu$.}  \label{tild-mu-cor}
\end{figure}
 
 \begin{equation}\label{2-lam1_corr}
\lambda_1^{<}=\lambda_1
\end{equation}

\begin{equation}\label{2-Dh_corr}
 D_h^{<}=D_h
\end{equation}

\begin{equation}\label{2-Dphi_corr}
 D_{\phi}^{<}=D_{\phi} + \frac{g_2^2 D_{\phi} D_h K_d}{\nu \tilde \mu^2 d}\int_{\Lambda/b}^{\Lambda}\frac{d q \, q^{d-1}}{q^2}
\end{equation}
 
 \begin{figure}[htb]
\includegraphics[width=8cm]{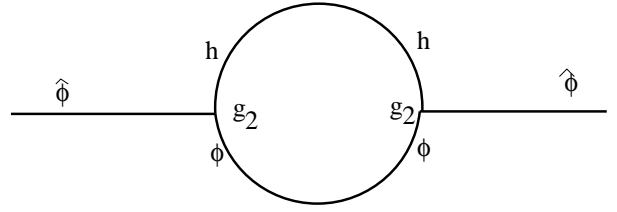}\hfill
\caption{One loop correction to $D_\phi$.}  \label{Dphi-cor-fig}
\end{figure}
 
 \begin{equation}\label{g2_cor}
  g_2^{<}=g_2 - \frac{g_2^3 D_h K_d}{\nu \tilde \mu^2 d}\int_{\Lambda/b}^{\Lambda}\frac{d q \, q^{d-1}}{q^2}
 \end{equation}

 \begin{figure}[htb]
\includegraphics[width=8cm]{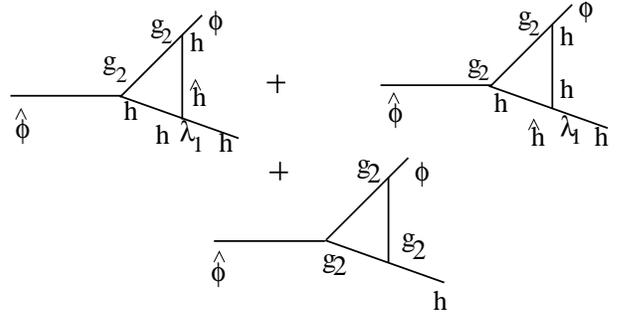}\hfill
\caption{One loop correction to $g_2$.}  \label{g2-cor}
\end{figure}
 
 The relevant Feynman diagrams for $\tilde\mu, 
 D_\phi$ and $g_2$ are given by Figs.~\ref{tild-mu-cor},\ref{Dphi-cor-fig} and 
\ref{g2-cor}, respectively.


\begin{thebibliography}{99}
\bibitem{comment1} Without the $\phi^4$-term in (\ref{free-en1}), it becomes 
admittedly artificial, but still suffices for our purposes here.
\end{thebibliography}
\end{document}